\documentclass[apl,twocolumn, amsmath, amssymb, superscriptaddress]{revtex4-1}

\usepackage{mathrsfs}
\usepackage{amsfonts}
\usepackage{graphicx}
\usepackage{dcolumn}
\usepackage{amsmath}
\usepackage{float}
\usepackage{epsfig}
\usepackage{booktabs}
\usepackage{subfigure}

\begin{document}

\title{First Photon Ghost Imaging}
\author{Xialin Liu}
\affiliation{State Key Laboratory of Advanced Optical Communication Systems and Networks, Shanghai Key Laboratory on Navigation and Location-based Service, and Center of Quantum Information Sensing and Processing(QSIP), Shanghai Jiao Tong University, Shanghai 200240, China}
%\email{lihuuhil@sjtu.edu.cn}
\author{Jianhong Shi}
\email{purewater@sjtu.edu.cn}
\affiliation{State Key Laboratory of Advanced Optical Communication Systems and Networks, Shanghai Key Laboratory on Navigation and Location-based Service, and Center of Quantum Information Sensing and Processing(QSIP), Shanghai Jiao Tong University, Shanghai 200240, China}
\author{Huichao Chen}
\affiliation{State Key Laboratory of Advanced Optical Communication Systems and Networks, Shanghai Key Laboratory on Navigation and Location-based Service, and Center of Quantum Information Sensing and Processing(QSIP), Shanghai Jiao Tong University, Shanghai 200240, China}
\author{Guihua Zeng}
\email{ghzeng@sjtu.edu.cn}
\affiliation{State Key Laboratory of Advanced Optical Communication Systems and Networks, Shanghai Key Laboratory on Navigation and Location-based Service, and Center of Quantum Information Sensing and Processing(QSIP), Shanghai Jiao Tong University, Shanghai 200240, China}%
\affiliation{College of Information Science and Technology,
Northwest University, Xi'an 710127, Shaanxi, China}
\begin{abstract}
Conventional imaging at low light level requires hundreds of detected photons per pixel to suppress the Poisson noise for accurate reflectivity inference. In this letter, we propose a high-efficiency photon-limited imaging technique, called first-photon ghost imaging, which recovers image from the first-photon detection by exploiting the physics of low-flux measurements and the framework of ghost imaging. The experimental results demonstrated that it could retrieve an image by only 0.1 photon detection per pixel, which is three orders lower than the conventional imaging technique. The SNR model of the system has been established for noise analysing. Our technique is supposed to have applications in many fields, ranging from biological microscopy to remote sensing.
\end{abstract}


\maketitle

\section{Introduction}

Photon-limited imaging has attracted great interest for its importance of application under extreme environment, such as night vision\cite{kallhammer2006imaging}, biological imaging\cite{Shen2000Two,McClatchy:16,becker2004fluorescence}, remote sensing\cite{mccarthy2013kilometer,molero2012anomaly}, spectral imaging in astronomy\cite{borkowski2013supernova}, etc., when off-the-shelf methods fail on photon limited data. Conventionally, the transverse spatial image is typically recovered by either a spatially resolving detector array with a floodlight illumination, or a single detector point by point by using raster-scanned illumination. However, even with time-resolved single-photon detectors, it still requires hundreds of photon per pixel to suppress Poisson noise inherent in photon counting for obtaining accurate intensity. In this way, extremely long time as well as much laser power will be expended for detection at low light level. And that may result in failure of imaging, such as biological imaging when samples could be destroyed by laser power or target tracing\cite{Magana2013Compressive} when the object moves very quickly.

At extremely low photon fluxes, photon detections are arrivals in a merged Poisson process of signal photon detections, background photon detections and dark counts\cite{migdall2013single}. To suppress the Poisson noise and have an accurate inference, the corresponding methods have been mainly carried through two ways: one is improving the Poisson intensity estimation model, and another is to design better measurement systems. Many earlier approaches for modeling and estimating Poisson process were based on wavelet-based methods\cite{nowak1997optimal,kolaczyk1999wavelet}. Multiscale photon limited imaging was first proposed by Timmermann and Nowak\cite{timmermann1997multiscale} in 1997, and has been improved and perfected after that\cite{timmermann1999multiscale,Willett2003Platelets,kolaczyk2004multiscale}. %That is an approach that partition image into multiscale homogeneous regions by using Poisson likelihood to determine optimal partition.
From this, more general and effective sparsity models for image have been discussed to deblur the Poisson noise, such as Toal variation and logarithmic regularization\cite{wright2009sparse,oliveira2009adaptive}, as well as sparsity models based on image patches\cite{makitalo2011optimal,salmon2014poisson} which underlie nonlocal mean, BM3D\cite{dabov2007BM3Dimage}, and dictionary learning. Kirmani, et al.\cite{kirmani2014first} proposed first-photon imaging(FPI) technique, which recovered image from first detected photon at each pixel.%, i.e., the photon efficiency achieved one detected photon per image pixel, denoted by 1 PPP.
They exploited spatial correlations to accurately reconstruct scene reflectivity by maximizing the product of data likelihoods over all spatial locations combined with a sparsity-promoting regularization function\cite{harmany2012spiral}. Additionally, the optimized design of the measurement system, such as compressive optical system\cite{duarte2008single,Ke2016Fast}, can facilitate the sensing capabilities. By exploiting techniques of compressed sensing and ghost imaging(GI) configuration\cite{pittman1995optical,Katz2009Compressive,Aspden:15}, Peter et al. obtained images from raw data comprised of fewer than one detected photon per pixel(PPP) by using entanglement source\cite{morris2015ghostimaging}. Considering the entanglement source is difficult to generate and transmit, with Time-Correlated Single-Photon-Counting(TCSPC) technique\cite{becker2005advanced}, Zeng's group implemented computational ghost imaging(CGI) at low light level by classical source\cite{yang2015computational}. Edgar, et al. realized 3D imaging with a single-pixel camera \cite{edgar2016first} by using the histograms of the arrival times of the first backscattered photon for each illumination pulse of each illumination pattern. That method requires at least hundreds of detected photons per image pixel which result in a lot of time expense.

Here, we theoretically and experimentally investigate a novel photon-limited imaging technique, the first-photon ghost imaging(FPGI)\cite{Liu:17}, which exploits the physics of low-flux measurements and the TCSPC-CGI configuration. It retrieves image efficiently from only first-photon data of each illumination pattern, and as the undersampling manner of GI, it can image with no more than 1 PPP. Further, exploiting the concept of time-correspondence differential ghost imaging\cite{Luo2012Nonlocal,li2013time}, a Fast First-photon ghost imaging(FFPGI) method has been raised with noticeable time saving. The experiment results show our scheme can reconstruct the object with PSNR of around 3dB by 0.1 PPP detection. The signal-to-noise ratio(SNR) model of our scheme has been established for noise analysing. The sparsity of premodulated patterns, as a vital influencing factor to the SNR, also has been discussed in this letter. Our technique has superiority at searching small target in big background with high efficiency and accuracy, and it facilitates the practical applications of ghost imaging ranging from biological microscopy to remote sensing.

\section{Imaging scheme and Noise model}

\begin{figure}[htbp]
\centering
\includegraphics[width=1.0\linewidth]{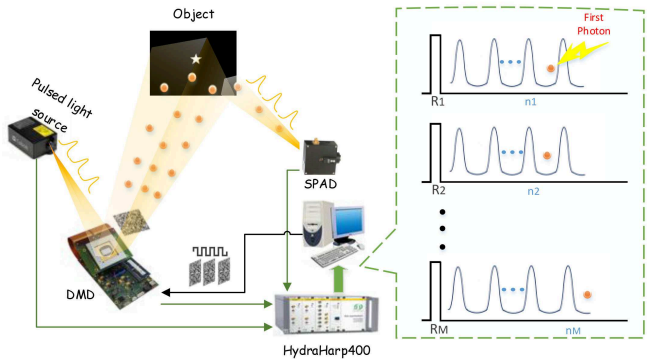}
\caption{The scheme of FPGI. The DMD modulates the spatial intensity of pulsed light source with binary random speckle patterns. The reflected light from the object was detected by a single photon avalanche diodes. The dashed box shows the time sequence of synchronous signals recorded by HydraHarp400.}
\label{fig:scheme}
\end{figure}

The imaging schematic is shown in Fig. \ref{fig:scheme}. A super-continuum pulsed laser with 1MHz repetition rate, is irradiated onto the programmable patterns of digital micromirror device(DMD), and then illuminating the object. DMD is an array of micromirrors consisting of 1024x768 independent addressable units, and each unit is a $13.6\mu m \times13.6\mu m$ micromirror with an adjustable angle of $\pm12^{\circ}$. At set intervals(eg.10ms), the DMD controller loads each memory cell with value '1' or '0', respectively representing $+12^{\circ}$ or $-12^{\circ}$ units which lead to the illuminated or non-illuminated pixels at the object plane. In our experiment, we use binary random speckle patterns, $R$, with $96\times128$ pixels where each pixel is constituted by $8\times8$ micromirror units. The sparsity of these patterns, i.e. the proportion of random '1' among all pixels, is adjustable. For every pattern, the corresponding first photon reflected from the object is recorded by the Single Photon Avalanche Diodes(SPAD)\cite{hadfield2009single}, and then the digital signal is input into the TCSPC module, along with the synchronous signals from the DMD and the laser pulses. So we can record the number of the pulses,${{{n}}_{{i}}}$, before the arrival of the first photon in $i$th sample pattern, ${{{R}}_{{i}}}$. The recorded first-photon data is used to estimate the intensity fluctuation for different patterns modulation.
\subsection{Reconstruction}

According to the characteristics of low flux measurements, the individual photon detection satisfies the Poisson Process. Let ${{{S}}_{{i}}}$ be the average number of laser photons arriving at the SPAD detector in response to a single-pulse illumination, $B$ denote the arrival rate of background photons at the detector, ${{{T}}_{{r}}}$ be the pulse repetition period, and $\eta$ be the efficiency of photodetection. Then the probability of no photon detected by a single pulse illumination is ${P_0}({{{S}}_{\rm{i}}}) = {e^{ - \eta ({S_{\rm{i}}}+B{T_{\rm{r}}})}}$. Because each pulse is independent, the number of pulses before first detection, denoted by $n$, has the geometric distribution,

\begin{eqnarray}
%\begin{aligned}
{P_{\rm{r}}}\left[ {{n_{\rm{i}}} = k} \right] = {P_0}{({{{S}}_{\rm{i}}})^{k - 1}}\left[ {1 - {P_0}({{{S}}_{\rm{i}}})} \right].
% = {e^{ - \eta (k - 1)({S_{\rm{i}}}+B{T_{\rm{r}}})}}(1 - {e^{ - \eta ({S_{\rm{i}}}+B{T_{\rm{r}}})}})
%\end{aligned}
\end{eqnarray}
In the absence of background, the pointwise maximum-likelihood intensity estimate,
 ${\mathop {{{{S}}_{\rm{i}}}}\limits^ \wedge  _{ML}}$, is proportional to ${{\rm{1}} \mathord{\left/
 {\vphantom {{\rm{1}} {{n_i}}}} \right. \kern-\nulldelimiterspace} {{n_i}}}$ for ${n_i} \gg 1$\cite{kirmani2014first}:
\begin{eqnarray}
%\begin{array}{l}
{{\hat S}_{ML}} = \mathop {\arg \max }\limits_{S > 0} \log \{ {e^{ - \eta ({n_{\rm{i}}} - 1)S}}(1 - {e^{\eta S}})\} \propto \frac{1}{{{n_{\rm{i}}}}}.
%\end{array}
\end{eqnarray}
Assuming that the reflection function of the object is ${O(x,y)}$, the total intensity ${{{S}}_{{i}}}$ could be described as
 \begin{eqnarray}
{{{S}}_{{i}}}{\rm{ = }}\iint{{R_i}(x,y)O(x,y)dxdy }.
\label{eq:Si}
 \end{eqnarray}

Thus the object could be retrieved by the correlation arithmetic with total iteration time of M,
\begin{eqnarray}
O_{{FPGI}}(x,y) = \frac{1}{M}\sum\limits_{i = 1}^M {\frac{1}{{{n_i}}}} ({R_i}(x,y) - \overline {{R_i}(x,y)} ).
\label{eq:fpgi}
\end{eqnarray}

Exploiting threshold technique in Time-correspondence differential ghost imaging\cite{li2013time}, we demonstrate a improved time-saving imaging method, FFPGI. The threshold $\bar{n}$ is set to select the efficient reference frames. If the first photon arrives before the $\bar{n}$th pulse of a certain pattern, we identify this pattern is effective, otherwise discard that pattern. And the iteration of the effective K patterns contributes to image reconstruction:
%statistical result of detection value as the
\begin{eqnarray}
O_{FFPGI}(x,y) = \frac{1}{K}\sum\limits_{j = 1}^K ({R_{j}}(x,y) - \overline {{R_{j}}(x,y)} )
\label{eq:ffpgi}
\end{eqnarray}

\begin{figure}[H]
\centering
\includegraphics[width=0.7\linewidth]{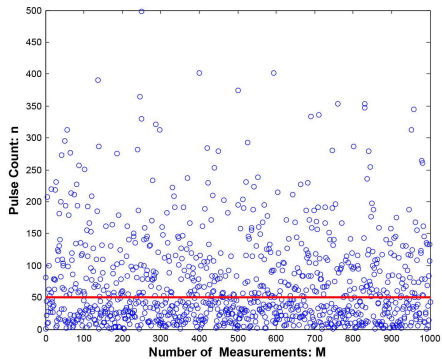}
\caption{Pulse count number of the first photons in 1000 Measurements. Data below the red line correspond to the patterns selected for FFPGI.}
\label{fig:ffpgi}
\end{figure}
Fig. \ref{fig:ffpgi} shows pulse counting graph of first-photon arrivals in 1000 measurements, where we use ${\bar{n}=50}$ as the threshold to select the patterns. The photon counting rate in our experiment is 0.83\%, i.e., there are 120 pulses in average before the first photon arriving. In practical, there is no need to detect the first photon which beyond the threshold. So, this method can save most time on the basis of FPGI.

\subsection{Noise Model}

The spatial structured light modulation is an essential step of our scheme. For a certain pattern illumination, we obtain the total of multi pixels' reflectivity by a single detection. And the number of pixel to be detected in a single measurement just is the number of pixels in '1' state on DMD, which is depend on the sparsity of the modulation patterns, denoted by $Sp$. For we use random binary patterns, the more pixels are illuminated by a single pattern means the more statistic noise in a single time detection. Theoretically, the decrease of the sparsity of the modulated patterns could increase the accuracy of intensity estimation. Let $\Delta F$ be the average amplitude fluctuation of a single pixel, and for an $N$ pixels object retrieve, the statistic fluctuation noise $\Delta S_{i,F}$ in each measurement can be described as:

\begin{eqnarray}
\Delta {S_{i,F}} = {\left( {Sp \cdot N \cdot \Delta F} \right)^2}.
\end{eqnarray}

While, considering the white noise of environment in real scene and the condition of single-pixel detection, the background noise would become primary factor instead of stochastic noise when reducing the sparsity of pattern to a certain degree, and the signal might be submerged. So let $\Delta b(x,y)$ be the white noise at each pixel, $\Delta B$ be the average of $\Delta b(x,y)$. The background noise for a single pattern, denoted by $\Delta S_{i,B}$, is as bellow,

\begin{eqnarray}
\Delta {S_{i,B}} = \sum\limits_x {\sum\limits_y {\Delta b(x,y)} }  = N \cdot \Delta B.
\end{eqnarray}

Therefore, for each measurement, the SNR of the imaging system is given by,

\begin{eqnarray}
\begin{aligned}
SNR = 10\lg{\frac{{\sum\limits_x {\sum\limits_y {{R_i}(x,y)O(x,y)} } }}{{\Delta {S_{i,F}} + \Delta {S_{i,B}}}}}\\
 = 10\lg\frac{{Sp \cdot N \cdot P}}{{{{\left( {Sp \cdot N \cdot \Delta F} \right)}^2} + N \cdot \Delta B}}\\
 = 10\lg\frac{P}{{Sp \cdot N \cdot \Delta {F^2} + \Delta B/Sp}},
 \label{eq:snr}
\end{aligned}
\end{eqnarray}
%_{10}o_{10}
where $P$ is the percentage of high-reflectivity pixels of the object. From the Eq.(8), we know that the trade off of statistic noise and background noise caused the peak of the curve, as the former increase with the sparsity of patterns increasing but the latter decrease.

\section{Experimental Results and Discussion}

\begin{figure}[htbp]
 \centering
  \subfigure[Experiment and simulation Results]{
    \label{exp:sparsity} %% label for second subfigure
    \includegraphics[width=0.4\textwidth]{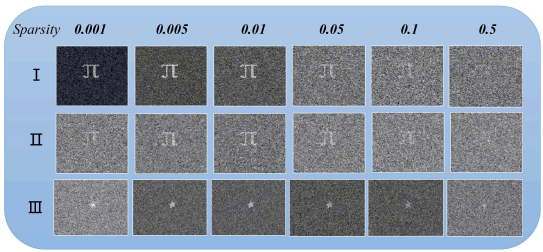}}

  \centering
  \subfigure[SNR vs. Sparsity]{
    \label{curve:snr} %% label for first subfigure
    \includegraphics[width=0.26\textwidth]{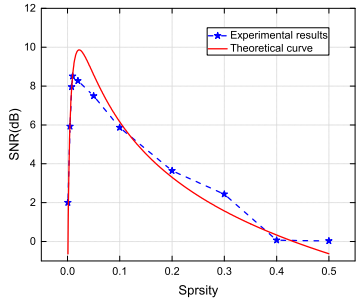}}      %% label for entire figure\label{fig:subfig}
  \subfigure[PSNR vs. Sparsity ]{
    \label{curve:psnr} %% label for first subfigure
    \includegraphics[width=0.26\textwidth]{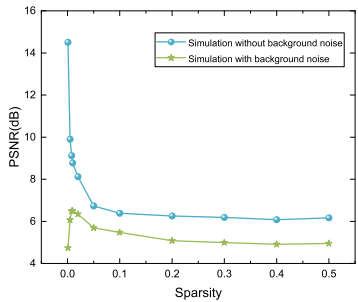}}      %% label for entire figure\label{fig:subfig}psnr_sprs
 \caption{Experimental and Simulation results with variable sparsity of the patterns from 0.001 to 0.5. (a) I,II: Simulation reconstruction of $100\times100$ object by 10000 measurements without(I)/ with(II) background noise. III: Experimental reconstruction of $96\times128$ object by 10000 measurements; (b) Theoretical SNR curve(red line) by Eq.\ref{eq:snr} and experimental SNR results(blue starred line). The parameter values of $P$, $\Delta F$ and $\Delta B$ in Eq.\ref{eq:snr} respectively are $10^{-2}, 1.5\times10^{-3}, 10^{-4}$; (c) The PSNR curve versus sparsity variation by simulation.} \label{exp:sparsity and snr}
 \hspace{0.1in}
 \end{figure}

In experiment, we adjust the sparsity from 0.001 to 0.5, and compare the results of simulation and experiment respectively(Fig.\ref{exp:sparsity and snr}). The simulation results without background noise show the lower sparsity the better reconstruction. While the results from experiment and simulation with background noise demonstrate too low sparsity also can reduce the reconstruction quality. The theoretical SNR curve and the experimental SNR versus sparsity of modulated patterns can be seen in Fig.\ref{curve:snr} which meet well one another. The combined effect of statistic noise and background noise also can be certified from the simulation of peak signal-to-noise ratio(PSNR) curve graph in Fig.\ref{curve:psnr} The PSNR is decreasing with the increasing sparsity by simulation without the background noise(Fig.\ref{curve:psnr} blue dot line), but there is a PSNR peak by simulation with background noise(Fig.\ref{curve:psnr} green starred line). That optimal sparsity value is the result of the trade-off of the statistic noise from modulation and the background white noise, which agree well with the theoretical model as Eq.\ref{eq:snr}. According to the above results, we choose the sparsity of 0.01 for modulation patterns in the following experiment.

\begin{figure}[htbp]
  \centering
  \subfigure[FPGI result.]{
    \label{FPGI} %% label for first subfigure
    \includegraphics[width=0.2\textwidth]{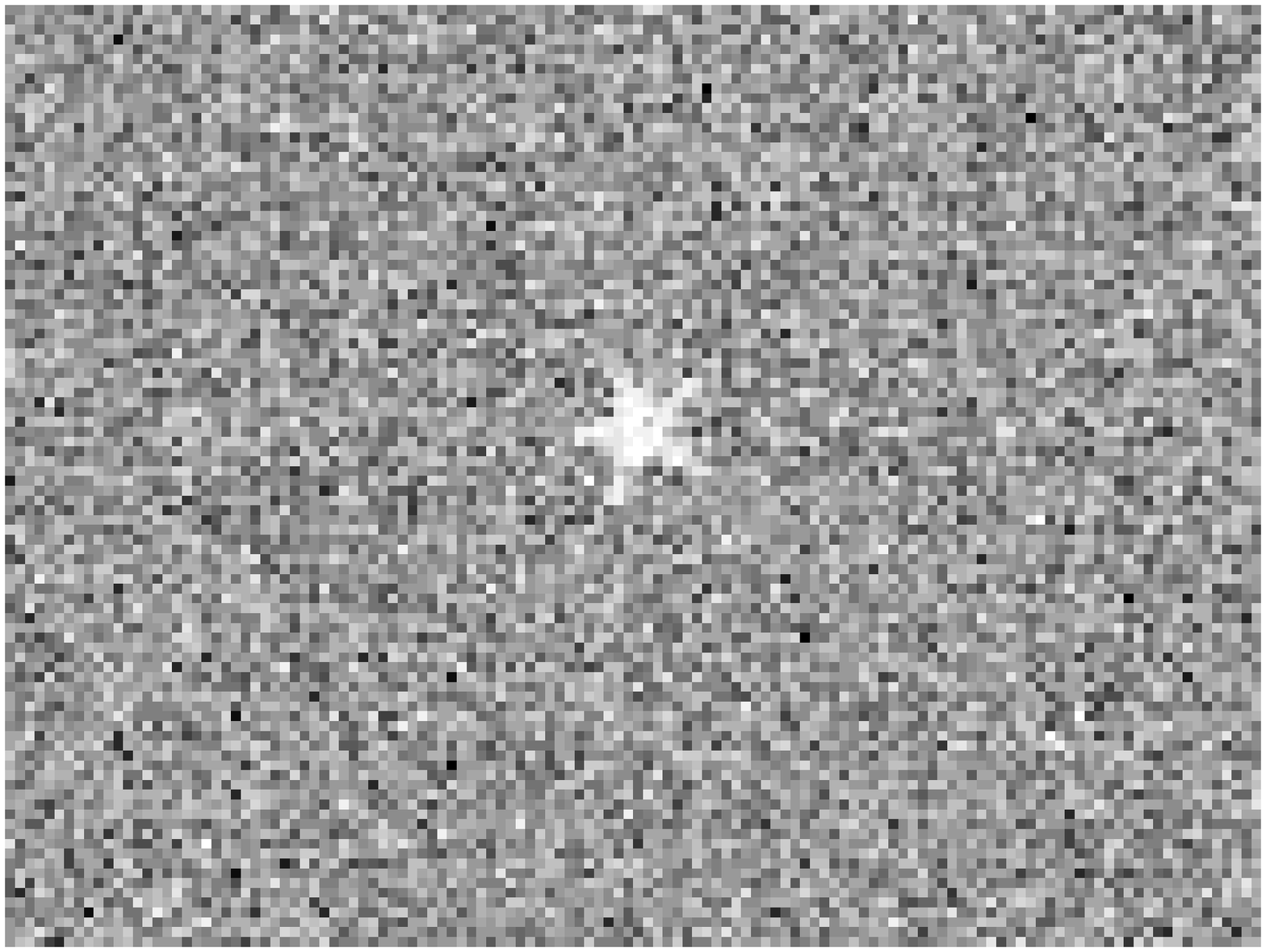}}
  \hspace{0.1in}
  \subfigure[After Roadfilter]{
    \label{roadfilter result} %% label for second subfigure
    \includegraphics[width=0.2\textwidth]{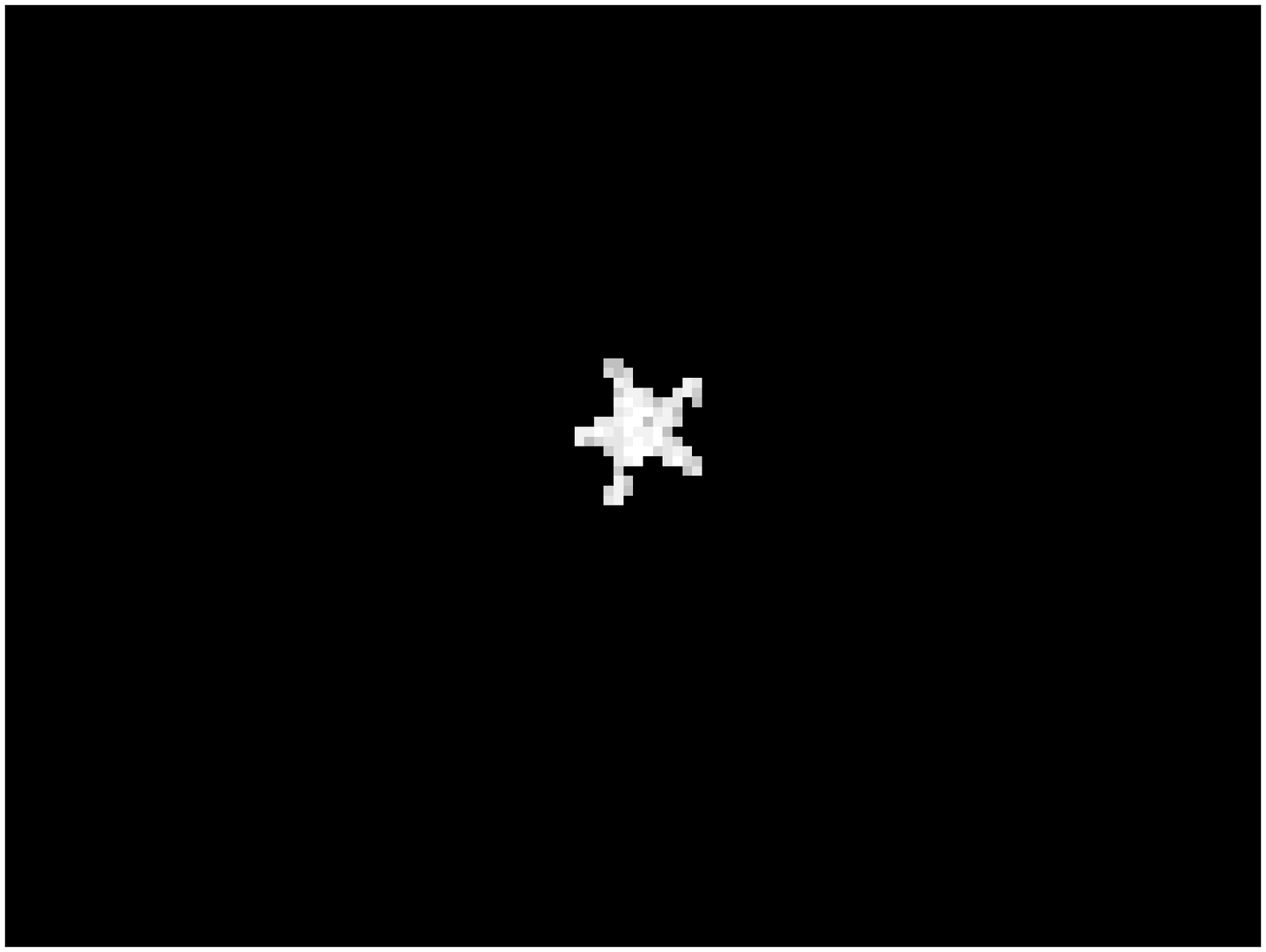}}
  \caption{Experimental result. Reconstruction of a 96x128 image with 10000 first-photon data.}
  \label{exp:roadfilter} %% label for entire figure
\end{figure}

To avoid the noise photons to be detected, we can set the time gate to the photons' time of flight(TOF) to filter out the noise photons which are not reflected from the object plane. That is because that the TOF of photons represents the different distance of the light sources plane to detection plane. What's more, the Roadfilter algorithm\cite{kirmani2014first} can be used to deblur the background noise and enhance the visibility of the image after the origin reflectivity estimation finished. It exploits the natural spatial correlation of the objects by computing the rank-ordered absolute differences statistic of the certain spatial location and its eight neighboring pixels' reflectivity, and then using the threshold to identify whether the photon detection was due to signal or noise. The threshold is dependent on the origin estimation from Eq.\ref{eq:Si} and Eq.\ref{eq:fpgi}. Fig.\ref{exp:roadfilter} shows the original reconstruction by FPGI from 10000 first-photon data and the result after the Roadfilter post-treatment. By iterating the Roadfilter several times, almost all background noise can be eliminated(Fig.\ref{roadfilter result}).

\begin{figure}[htbp]
\centering
  \subfigure[A: FPGI results; B: FFPGI results.]{
    \label{ppp_pulse} %% label for first subfigure
    \includegraphics[width=0.44\textwidth]{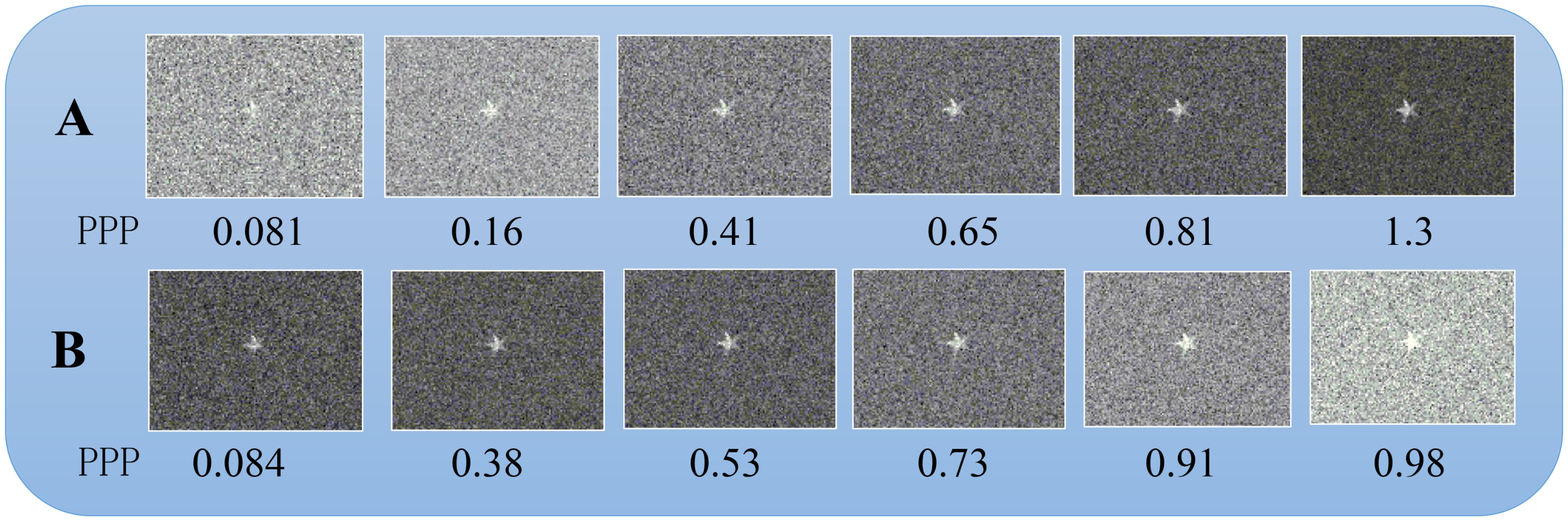}}
  \subfigure[PSNR of FPGI]{
    \label{psnr:ppp} %% label for first subfigure
    \includegraphics[width=0.2\textwidth]{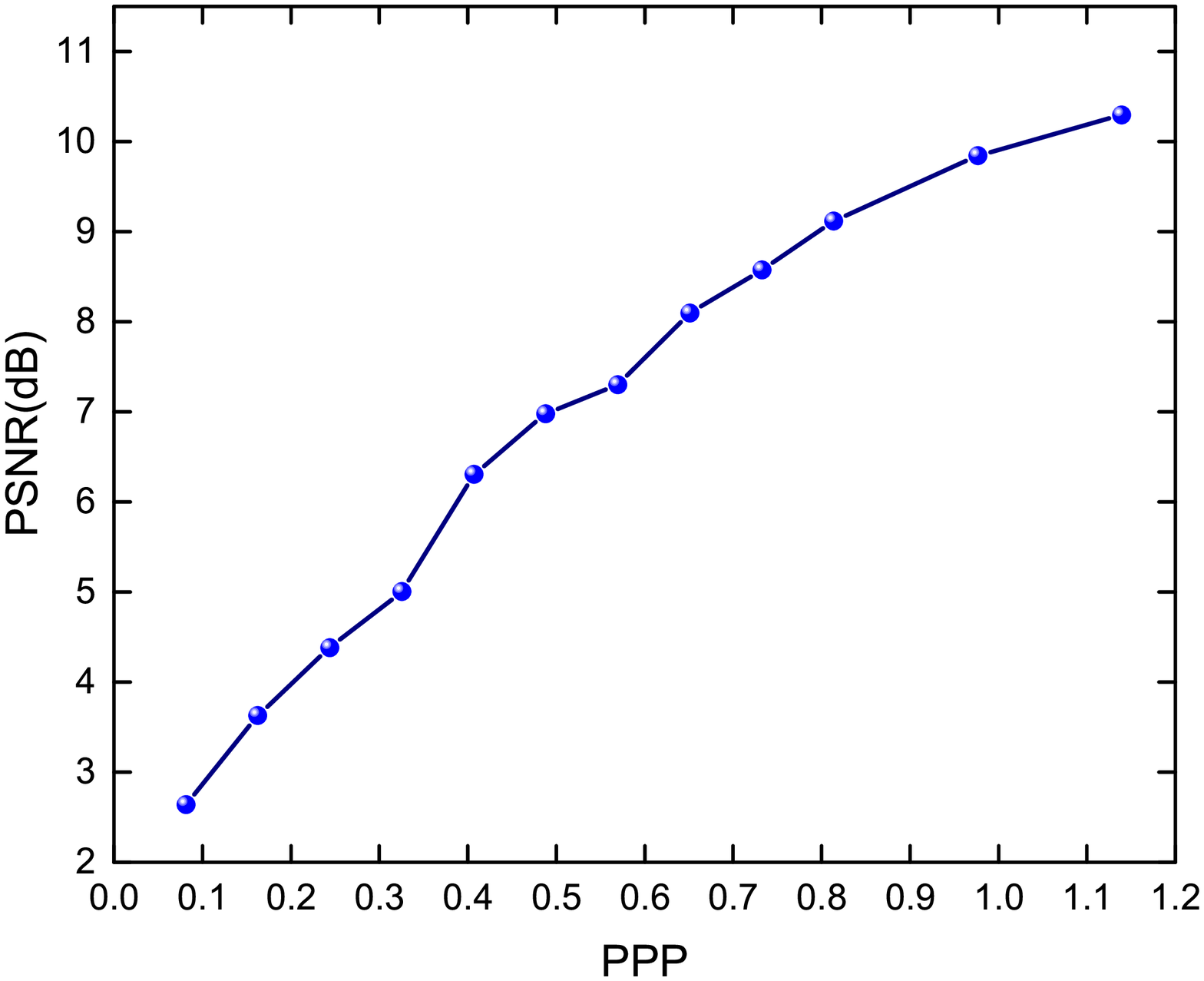}}
  \hspace{0.1in}
  \subfigure[PSNR of FFPGI]{
    \label{psnr:pulse} %% label for second subfigure
    \includegraphics[width=0.2\textwidth]{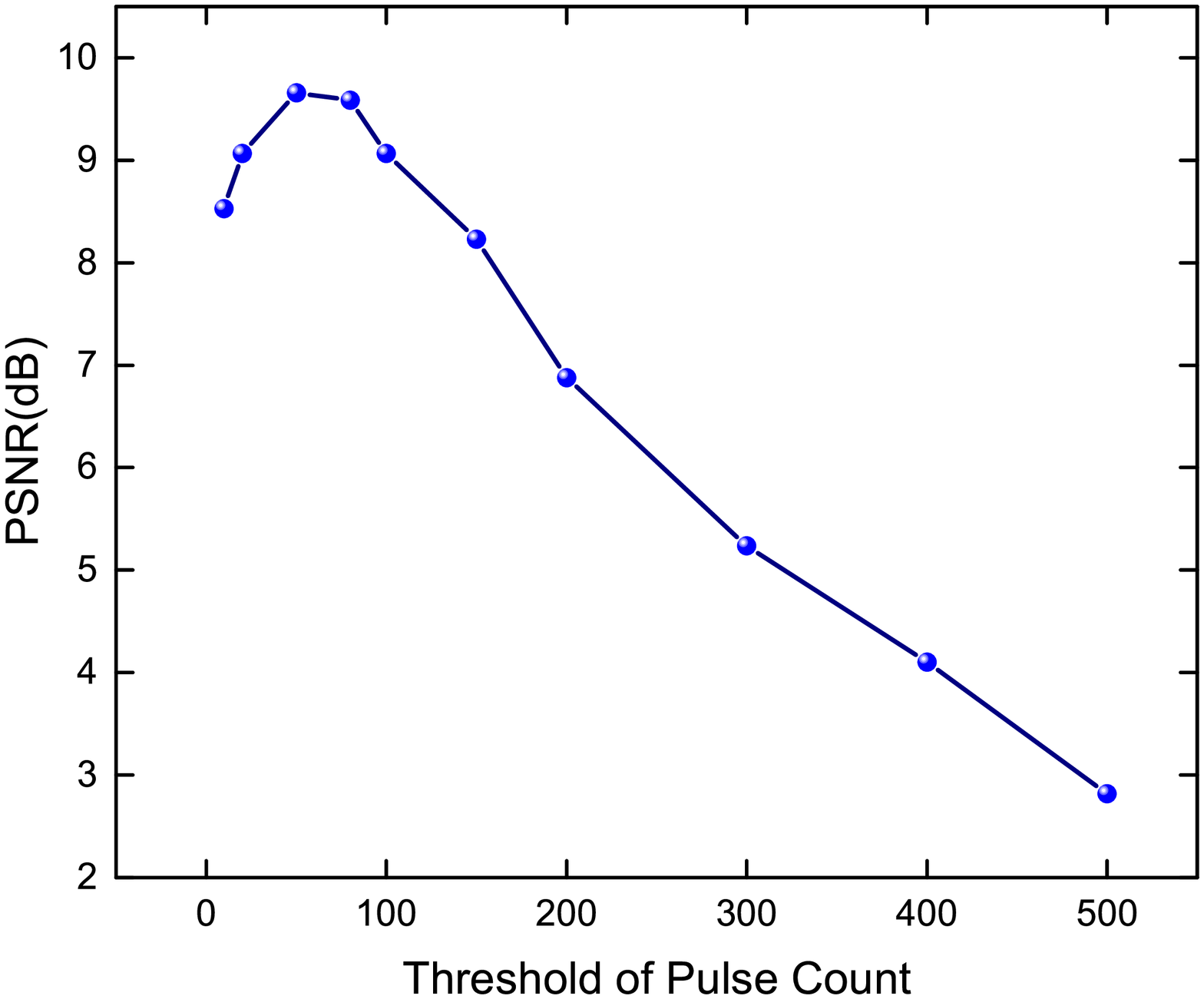}}
  \caption{Experimental result. The FPGI results in (a)A with different PPP are operated by adjust the measurement times from 1000 to 10000; and (a)B is the FFPGI results with the variable pulse-count thresholds from 10 to 500 by 10000 measurements.}
  \label{FPGI,FFPGI} %% label for entire figure
\end{figure}

Fig.\ref{ppp_pulse} demonstrates the results of FPGI and FFPGI. As we can see, the PPP could be fewer than 0.1, that is, reconstructing a 96$\times$128 pixels picture just by using the information from 1000 first-photon data. And with the improving PPP, the details of the object become more distinct(Fig.\ref{ppp_pulse} line A). It also can be seen clearly from the PSNR curves in Fig.\ref{psnr:ppp}. While the FFPGI results in Fig.\ref{ppp_pulse} line B. show that as the PPP increasing, the background noise become more evident. That is because the more detected first photons(i.e. the higher pulse threshold value), the more patterns are identified to be overlapped, and meanwhile the more background noise join. With the proper pulse number threshold value(Fig.\ref{psnr:pulse}), the clear image can be obtained by this unusually terse method. Furthermore, it is also worth attention that the less pulse number as threshold, accordingly the less time it will cost for image.

\section{Conclusion}

The comparison of CGI, FPI, FPGI and FFPGI are shown in Table. \ref{tab: compaison}. The time is only calculated for signal detection, which neglects the scanner scanning time and computer running time. As we can see, ghost imaging technique needs less measurements than FPI, and it only requires fixed single-pixel detector without spatial resolving. FPGI and especially FFPGI have higher photon efficiency than the previous two techniques, just as remarkable low PPP and time cost.
%The numerical value is calculated on the condition of reconstructing a 96x128 pixels image at 0.83\% photon counting rate light level and 1MHz laser pulse repetition rate. several fundamental characteristics among
\begin{table}[htbp]
\centering
\caption{\bf Comparison of characteristics of CGI, FPI, FPGI, and FFPGI to reconstruct a 96x128 pixels image at low light level of 0.83\% photon counting rate.  }
\begin{tabular}{ccccc}
\hline
\hline
Condition & PPP & Measurements & Time/s & Detector\\
\hline
CGI & $10^2-10^3$ & 2000 & 10 & single-pixel\\
%\hline
FPI & $\geq1$ & 12288 & 1.5 & scanner\\
%\hline
FPGI & 0.16 & 2000 & 0.24 & SPAD\\
%\hline
FFPGI & 0.08 & 2000 & 0.02 & SPAD\\
\hline
\hline
\end{tabular}
\label{tab: compaison}
\end{table}

In conclusion, the FPGI technique can achieve high-efficiency performance at extremely low light level. By using first-photon data for intensity estimation and correlation imaging framework for spatial reconstruction, the experimental and simulation results show our scheme can reconstruct the object with only 0.1 photon detection per image pixel. Considering the characteristic of correlation imaging, the SNR model has been established for analysing the influence of the sparsity of modulated patterns to system noise. Our technique can extract more information from the collection of single detection than current imaging methods. Thus, it saves a lot time as well as laser power, which can be crucial for biological applications, such as fluorescence-lifetime imaging. It also is superior at remote sensing, such as recognizing the small object in wide field view with high efficiency and accuracy. This method may be applied to enhance a variety of computational imagers that rely on sequential correlation measurements.
%make itself generally practical in photon-limited condition, such as night vision, spectral imaging in astronomy Especially,achieve higher quality image with even lower PPP
The FPGI system we have demonstrated can be improved by exploiting compressed sensing technique as well as superior sparsity model for intensity estimation. What's more, by utilizing the arrival-time data of the first photon for range dimension sensing, this scheme can also be used for 3D imaging.

\section*{Funding Information}
National Natural Science Foundation of China(NSFC) (Grants No: 61471239, 61631014); Hi-Tech Research and Development Program of China (2013AA122901).
%\section*{Acknowledgments}

% Bibliography
\bibliography{myReference}
\end{document}